\documentclass[fleqn,10pt]{wlscirep}
\usepackage[utf8]{inputenc}
\usepackage[T1]{fontenc}
\usepackage{textcomp}
\usepackage{array}
\usepackage{listings}
\usepackage{setspace}
\usepackage{mathptmx}
\usepackage{colortbl}
\usepackage{graphicx}
\usepackage{amssymb, amsmath, amsfonts}
\usepackage{amsthm}
\usepackage{subfig}
\usepackage{epsfig}
\usepackage{times}
\usepackage{float}
\usepackage{rotating}
\usepackage{makeidx}
\usepackage{url}
\usepackage{multirow}
\usepackage{booktabs}
\usepackage{tabularx}
\usepackage{varwidth}
\usepackage{pdflscape}
\usepackage{placeins}
\usepackage{threeparttable}
\usepackage{adjustbox}
\usepackage[intoc]{nomencl}

\usepackage[subfigure, titles]{tocloft}
\usepackage{datetime}
\usepackage{siunitx}

\usepackage{algorithm}
\usepackage{algorithmicx}
\usepackage{algpseudocode}
\usepackage{lineno}

\definecolor{darkblue}{RGB}{0,0,170}
\newcommand{\hl}[1]{{#1}}
\newcommand{\celsius}{\SI{}{\degreeCelsius}}

\title{Immersive and Wearable Thermal Rendering for Augmented Reality}

\author[1,*]{Alexandra Watkins}
\author[2]{Ritam Ghosh}
\author[1]{Evan Chow}
\author[1,2]{Nilanjan Sarkar}
\affil[1]{Vanderbilt University, Mechanical Engineering, Nashville, 37212, USA}
\affil[2]{Vanderbilt University, Electrical and Computer Engineering, Nashville, 37212, USA}

\affil[*]{alexandra.watkins@vanderbilt.edu}

\begin{abstract}
\hl{We present a proof-of-concept palm-mounted thermal feedback prototype addressing thermal rendering challenges specific to augmented reality (AR), where users must interact with both real and virtual objects in their physical workspace. In contrast to thermal feedback systems developed for virtual reality, AR thermal feedback must preserve manual dexterity, maintain access to real-world thermal cues, and provide coherent virtual temperature sensations without obstructing natural object interaction. We propose three AR-specific design considerations, which our prototype implements: indirect feedback to preserve fingertip dexterity, active thermal passthrough to sense and render the temperature of contacted physical surfaces, and spatially and temporally varying thermal rendering across the palm. Human-subject experiments evaluated perceptual sensitivity, indirect feedback, active thermal passthrough, spatial pattern recognition, and moving thermal rendering during AR interaction. Results showed that although indirect feedback reduced perceived realism during visual contact at the fingertips, it did not reduce immersion or comfort; active thermal passthrough supported temperature discrimination between real and rendered surfaces; and spatiotemporal rendering significantly improved immersion and realism compared with static thermal stimulation. These findings suggest that our design considerations are viable design strategies for AR thermal haptics, while also clarifying tradeoffs for applications that require precise realism versus broader immersive thermal experience.}
\end{abstract}

\begin{document}

\flushbottom
\maketitle
%
%
\thispagestyle{empty}

\section{} \label{sec:Intro}

\hl{Thermal sensations are an important part of everyday object interaction. Warmth and coolness help people interpret material properties, environmental conditions, comfort, and safety, such as when holding a hot beverage or touching a cold metal rail. These experiences are mediated by thermoreceptors in the skin, which detect changes in temperature and support material discrimination and behavioral responses \cite{jonesMaterialDiscriminationThermal2003}. In immersive systems, thermal feedback can therefore contribute to presence, interactivity, and realism by making virtual or augmented objects feel more consistent with the physical experiences they represent \cite{peirisThermoVRExploringIntegrated2017,gibbsComparisonEffectsHaptic2022,lontscharAnalysisHapticFeedback2020}. As immersive technologies are increasingly used in education, workforce training, healthcare, and industrial processes \cite{johnReviewMixedReality2020,santiAugmentedRealityIndustry2021,bazavanVirtualRealityAugmented2021}, the need for effective multisensory feedback has also increased. Prior work has shown that the fidelity of immersive feedback can influence learning outcomes, task efficiency, and user satisfaction \cite{sirakayaAugmentedRealitySTEM2022,liberatoreVirtualMixedAugmented2021,akgunEffectsImmersiveVirtual2022,guerra-tamezImpactImmersionVirtual2023}. Thermal feedback is one important component of this broader challenge, particularly in AR, where virtual and augmented objects must be perceived in relation to the physical surfaces, materials, and interactions that surround them.}

\hl{AR introduces design constraints that differ from those of virtual reality (VR). AR systems overlay digital information onto the user's physical environment, allowing virtual content to be perceived and manipulated in relation to real objects and spaces \cite{milgramTaxonomyMixedReality1994,azumaSurveyAugmentedReality1997}. In contrast to fully virtual environments, AR users may need to grasp actual tools, touch real surfaces, and maintain awareness of physical material properties while simultaneously receiving feedback from virtual elements. Thermal feedback for AR therefore requires more than adding temperature cues to the hand; it requires strategies that balance physical interaction, real-world thermal perception, and coherent virtual sensation. Although recent AR head-mounted displays provide increasingly sophisticated visual and auditory feedback \cite{AppleVisionPro}, haptic and thermal feedback remain important for supporting the perceived presence of virtual objects and interactions \cite{garcia-valleEvaluationPresenceVirtual2017,krogmeierHumanVirtualCharacter2019,richardStudyingRoleHaptic2021,basdoganExperimentalStudyRole2000,slaterPresenceConsciousnessVirtual2005}.}

\hl{Compared with force, vibration, and tactile feedback, thermal feedback remains relatively underexplored in immersive interfaces \cite{peirisThermoVRExploringIntegrated2017}. Existing systems demonstrate that thermal stimuli can improve realism and presence \cite{jonesMaterialDiscriminationThermal2003,hoMaterialRecognitionBased2018,caiThermAirGlovePneumaticGlove2020}, and recent wearable or multimodal devices have incorporated thermal cues into gloves, rings, handheld devices, and actuator arrays \cite{sunAugmentedTactileperceptionHapticfeedback2022b,ohLiquidMetalBased2021,zhuHapticfeedbackSmartGlove2020,huangSkinintegratedMultimodalHaptic2023}. These approaches show the value of thermal feedback, but they also reveal a design conflict for effective use in AR. Devices that cover the fingers or require handheld controllers \cite{azizHapticHandshankHandheld2020} may interfere with grasping, manipulation, or natural thermal sensing of physical objects. Such compromises may be acceptable in fully virtual contexts, but they become more consequential when users must interact fluidly with both physical and virtual entities in the same workspace.}

\begin{figure}[htbp]
    \centering
    \includegraphics[width=\textwidth]{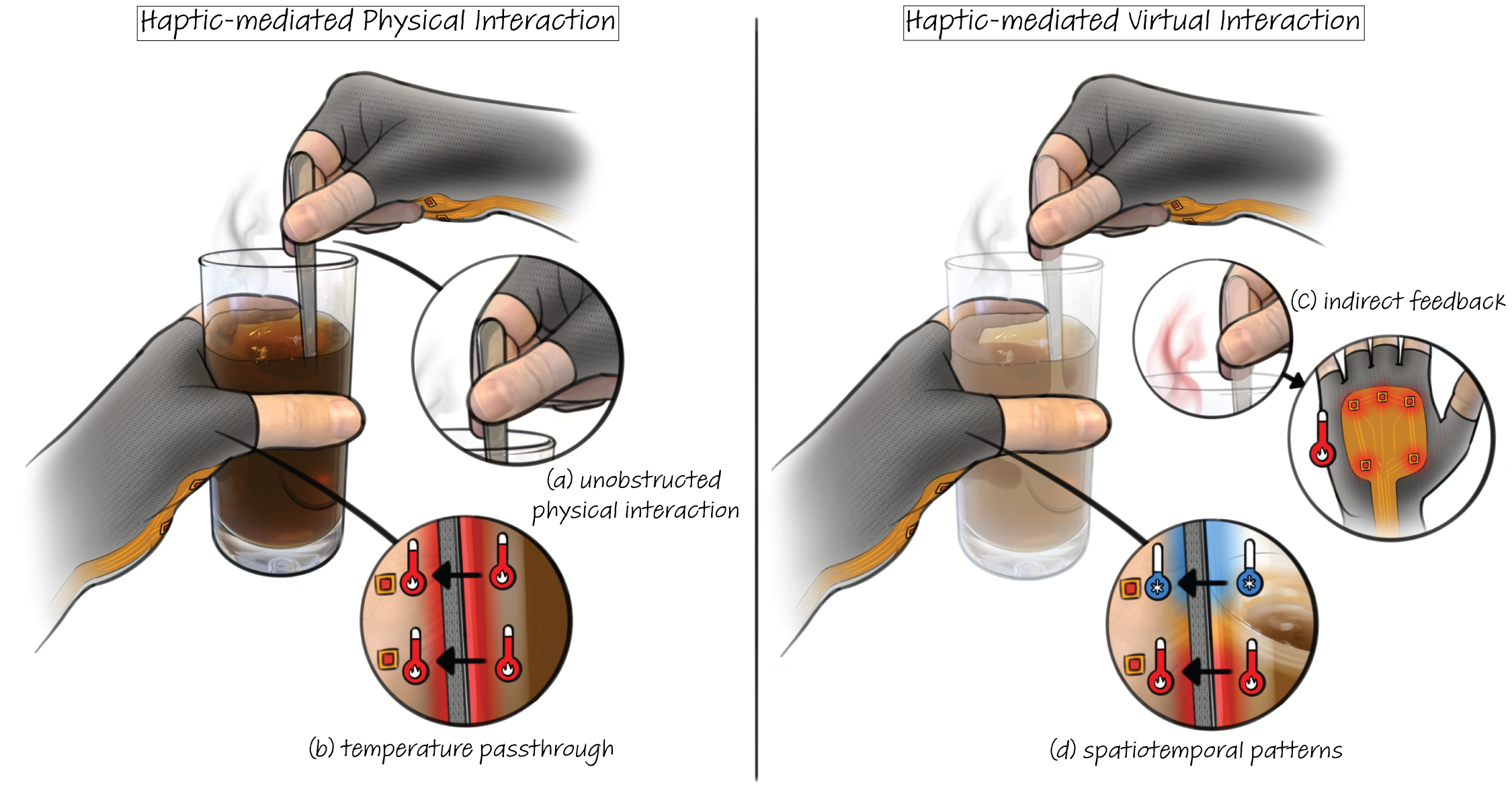}
    \caption{\hl{\textbf{Haptic-mediated thermal interactions in AR enabled by the proposed design considerations.}\\
    (\emph{a}) \emph{Physical interaction.} Leaving the fingers unobstructed supports fine manipulation of physical tools or objects and preserves direct thermal sensing at the fingertips. 
    (\emph{b}) \emph{Active thermal passthrough.} Temperature sensors on the external surface of a haptic device allow collocated thermal actuators on the interior surface to render the thermal properties of contacted surfaces. 
    (\emph{c}) \emph{Indirect feedback.} Thermal sensations applied to nearby occluded regions, such as the palm, can provide thermal cues during virtual interactions while avoiding actuation on the fingertips. 
    (\emph{d}) \emph{Spatiotemporal rendering.} Independently controlled and spatially arranged thermal actuators enable sensations that vary across the contact surface and over time, such as the walls of a virtual cup being covered and uncovered by warm liquid.}}
    \label{fig:Ch4-ConceptualOverview}
\end{figure}

\hl{We believe these AR-specific constraints motivate three design considerations for thermal feedback, as demonstrated in the use case of Figure \ref{fig:Ch4-ConceptualOverview}. First, indirect feedback can relocate thermal actuation away from the fingertips to reduce interference with manual dexterity. Second, active thermal passthrough can address the loss of natural thermal sensing caused by device occlusion by actively measuring the temperature of contacted surfaces and rendering that temperature to the skin. Third, spatiotemporal thermal rendering can support thermal sensations that vary across the contact surface and over time, allowing virtual thermal events to better correspond to dynamic AR interactions. Together, these considerations frame thermal feedback for AR as a problem of balancing real-world touch, virtual object feedback, and mixed physical--virtual coherence.}

\subsection{Present Work} \label{subsec:PresentWork}

\hl{In this manuscript, we present a novel proof-of-concept palm-mounted thermal-rendering prototype that implements these design considerations for AR (Figure \ref{fig:Ch4-DevicePics}). The palm provides sufficient surface area for multiple thermal actuators while leaving the fingertips available for grasping, manipulation, texture perception, and direct contact with real objects. This form factor allows the device to explore indirect thermal feedback, active thermal passthrough, and spatially distributed thermal rendering without fully occupying the contact surfaces most often used for precise manual interaction. The system combines independently controlled thermoelectric modules (TEMs) with outward-facing contact-detection and temperature sensors to provide warm and cool sensations, active thermal passthrough, and time- and space-varying temperature sensations rendering on the palm using only thermal actuators.}

\hl{We provide supporting evidence for our design considerations and their implementation in a proof-of-concept device using five controlled human-subject experiments. These experiments confirm the perceptual sensitivity to warm and cool stimuli on the palm as an appropriate location for haptic feedback, the user experience tradeoffs of indirect feedback, the ability of active thermal passthrough to support object temperature perception, recognition of spatial thermal patterns, and the effect of moving thermal rendering during AR interaction. This work contributes an AR-specific framing for thermal feedback, a novel wearable prototype implementation of three design considerations, and empirical evidence that these considerations can support thermal interaction while preserving the physical sensing and manual dexterity that distinguish AR from fully virtual environments. In particular, we demonstrate that the designed AR thermal interaction prototype for AR applications ensure immersion.}

\FloatBarrier

\section{Results} \label{sec:Results}

\subsection{Design considerations for AR thermal feedback} \label{subsec:DesignReqs}

Thermal feedback for AR must balance virtual object rendering with continued interaction in the physical environment (Figure \ref{fig:Ch4-ConceptualOverview}). Unlike fully virtual systems, AR interfaces require users to grasp, manipulate, and thermally sense real objects while also perceiving temperature cues associated with virtual content. \hl{This creates a set of design tradeoffs: thermal cues must be perceptible and expressive, but the mechanisms used to render them should not substantially obstruct manual dexterity, interfere with physical contact, or eliminate access to real-world thermal information. These constraints motivate three design considerations for AR thermal feedback:}

\begin{itemize}
    \item[\textbf{DC 1:}] \hl{Use indirect feedback to preserve manual dexterity.}
    \item[\textbf{DC 2:}] \hl{Use active thermal passthrough to preserve access to real-world temperature information.}
    \item[\textbf{DC 3:}] \hl{Use spatiotemporal rendering to support complex and dynamic virtual thermal interactions.}
\end{itemize}

The first two considerations address how thermal feedback can coexist with real-world touch, while the third addresses how virtual thermal cues can vary across space and time to better match AR object interactions.

\subsubsection{DC 1: Use indirect feedback to preserve manual dexterity}

\hl{A first consideration is whether thermal feedback can be delivered without occupying the fingertips, which are often needed for exploratory and manipulative touch in AR. Bhatia, Hornb\ae k, and Seifi \cite{bhatiaAugmentingFeelReal2024} identified four types of touch interactions commonly used in AR from the set of exploratory procedures previously described by Lederman and Klatzky \cite{ledermanHandMovementsWindow1987}. Three of these interaction types require the thumb or fingers. Directly mounting thermal actuators on these contact surfaces can therefore interfere with grasping, manipulation, and natural sensing of physical objects.}

\hl{Indirect feedback addresses this issue by applying thermal cues to a nearby skin region rather than the apparent or functional point of contact. Prior work has explored displaced thermal stimulation at locations such as the base or back of the finger \cite{sanzcozcolluelaGeneratingMultimodalTextures2026,wangFieryHandsDesigning2024,zhuSenseIceFire2019} and the wrist \cite{peirisThermalBraceletExploringThermal2019}. An additional candidate location explored by our prototype device is to use the palm as the indirect feedback site, as it provides a relatively large surface area for thermal actuation while leaving the fingertips available for precise physical interaction.}

\hl{This approach introduces an important perceptual tradeoff. Relocating feedback away from the apparent point of contact may preserve dexterity, but it may also reduce realism if the visual and thermal locations do not align. This tradeoff is especially relevant because reducing actuator size or moving actuators away from the fingers can also limit heat transfer, contact area, and perceived stimulus intensity \cite{manasrah_perceived_2017}. Thermal sensitivity varies across the body and depends on both absolute temperature and rate of temperature change \cite{jonesWarmCoolLarge2008,filingeriThermosensoryMicromappingWarm2018,caldwellMechanoThermoProprioceptor1997}. By strategically positioning smaller, low-power actuators on areas with a higher thermoreceptor density or greater thermal sensitivity -- the palm acting as a great example -- we believe designers can maintain a significant thermal effect without sacrificing the user’s dexterity or comfort.}

\subsubsection{DC 2: Use active thermal passthrough to preserve access to real-world temperature information}

\hl{The last type of common AR touch interaction identified by Bhatia et al. \cite{bhatiaAugmentingFeelReal2024} is static contact and holding, where users maintain sustained touch with an object or surface. This method of interaction is particularly important for perceiving thermal properties of the environment, as humans intuitively use various regions of their skin, such as the palms, fingers, or even forearms, to gather thermal cues. However, when a thermal feedback device that requires contact is placed on these sensitive regions, it blocks direct contact with the physical environment. This creates a challenge: the user is no longer able to feel the true thermal characteristics of the object they are holding, potentially diminishing their sense of immersion and realism. Simply utilizing indirect feedback by itself does not adequately address this, as there is still the possibility the area of skin occluded by the device is needed for intuitive interactions.}

\hl{We propose a solution that we term \emph{active thermal passthrough}: the device measures the temperature of a contacted physical surface using outward-facing sensors and renders a corresponding sensation to the skin using inward-facing actuators. This approach does not passively transmit the object's temperature through the device. Rather, it actively senses and recreates a device-mediated thermal cue. This introduces a tradeoff: active thermal passthrough can preserve access to real-world temperature information, but the rendered cue is necessarily mediated by sensor response, actuator dynamics, contact conditions, and the thermal properties of the device interface. Active thermal passthrough is therefore intended to reduce the loss of real-world thermal information introduced by the device itself, allowing users to retain access to contact-based temperature sensations while still benefiting from augmented thermal feedback.}

\subsubsection{DC 3: Use spatiotemporal rendering to support complex and dynamic virtual thermal interactions}

\hl{A third consideration is that virtual thermal cues may need to vary across both space and time. Real and virtual object interactions are rarely uniform: contact may occur over only part of the hand, the hand may move across a surface, or the thermal state of a virtual object may change during interaction. A single static thermal stimulus may therefore be insufficient for interactions in which the perceived contact location or temperature distribution changes over time.}

\hl{Spatially varying thermal cues can contribute to the perception of surface properties and localized contact \cite{hiraiThermalTactileSensation2019}. This need is consistent with exploratory procedures such as lateral motion and contour following, in which dynamic touch interactions reveal properties of textures, shapes, and surfaces \cite{ledermanHandMovementsWindow1987}. Prior systems have demonstrated promising approaches to spatial thermal feedback, including benchtop thermal tactile displays \cite{hiraiThermalTactileSensation2019} and wearable skin-like actuators \cite{leeStretchableSkinLikeCooling2020}. However, spatial thermal rendering remains less explored in compact interfaces intended to support natural AR object interaction.}

\hl{Temporal variation is also important when contact changes during motion. Existing approaches often combine a static thermal source with vibrotactile cues to create the illusion of moving heat or cold. This can rely on thermal referral, in which a stationary thermal stimulus appears to move when paired with shifting tactile stimulation \cite{greenLocalizationThermalSensation1977, wangFieryHandsDesigning2024, singhalThermalMotionDesigning2024}. Liu et al. \cite{liuThermoCaressWearableHaptic2021} used this approach by overlaying static thermal output with moving tactile feedback, while Son et al. \cite{sonUpperBodyThermal2023} developed a vest that combined vibrotactile and thermal actuators to render moving thermal sensations. Nakatani et al. \cite{nakataniNovelMultimodalTactile2018} similarly combined a low-resolution grid of thermoelectric modules with a higher-resolution vibrotactile matrix to create a sense of thermal motion.}

\subsection{Thermal Device Design}
\label{sec:Design}

\begin{figure}[htbp]
    \centering
    \includegraphics[width=0.8\textwidth]{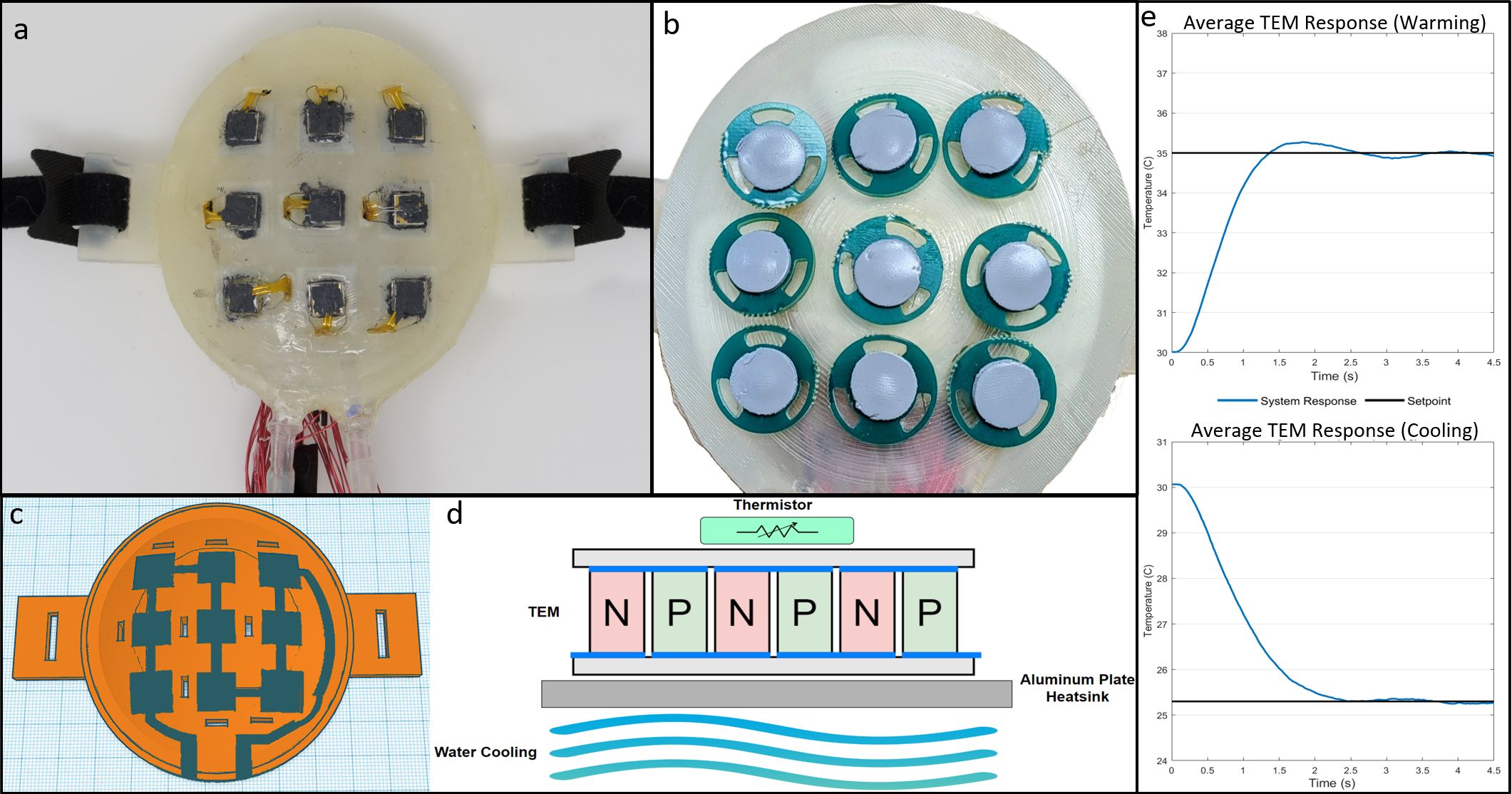}
    \caption{\textbf{Overview of the proof-of-concept wearable thermal feedback system.}\\
    (\emph{a}) Thermal feedback is applied by an array of individually controlled Peltier devices with temperature feedback from thermistors mounted to each contact surface. 
    \hl{(\emph{b}) Outward-facing thermistors covered by a thermally conductive pad allow measurement of contacted surface temperatures for active thermal passthrough. Reflective phototransistors mounted on the bottom of each pcb allow for contact detection.}
    (\emph{c}) Interior to the silicone base is a water-cooling channel for temperature management. 
    (\emph{d}) A single thermal actuation unit consists of a thermoelectric module and thermistor mounted on an aluminum base that acts as a water-cooled heatsink. 
    (\emph{e}) Average temperature response to a step input while in contact with a human palm at an ambient temperature of 30\celsius.}
    \label{fig:Ch4-DevicePics}
\end{figure}

We present a novel proof-of-concept wearable thermal feedback prototype designed to implement the three AR thermal feedback design considerations described above (Figure \ref{fig:Ch4-DevicePics}). The device consists of an array of TEMs arranged in a $3 \times 3$ grid against the palm. Each module functions as an individual heat pump and can be independently controlled, allowing the device to support spatially and temporally varying thermal cues without additional modalities of feedback as is typically done for thermal referral. A $10~\mathrm{k}\Omega$ thermistor was attached to the active surface of each TEM for closed-loop temperature control. Each TEM--thermistor assembly was mounted on a $10~\mathrm{mm} \times 10~\mathrm{mm}$ aluminum plate and embedded within a flexible silicone base. Internal water-cooling channels manage the temperature of the aluminum base plates, which act as heat sinks and reduce thermal buildup during sustained operation.

\hl{To support active thermal passthrough, the outward-facing surface of the device includes a $3 \times 3$ grid of custom PCBs with a surface contact sensor (VCNT2020 reflective phototransistor) and a $10~\mathrm{k}\Omega$ thermistor. Each thermistor is covered with a $0.2~\mathrm{mm}$ thermally conductive silicone pad to improve thermal coupling with contacted surfaces while protecting the sensor package. The phototransistor is mounted on the bottom of the custom PCB, and registers contact when enough pressure is applied to close the air-gap between the bottom of the pcb and the device's silicone body. Together, these outward-facing sensor modules allow the device to detect physical contact and measure contacted surface temperature, enabling active thermal passthrough when the palm is occluded by the device.}

The device was mounted against the user's palm to operationalize indirect feedback while preserving the fingertips for grasping, manipulation, and direct physical contact. The palm remains near common fingertip interaction sites, can itself serve as a direct contact surface for virtual objects, and provides a relatively large area for rendering thermal sensations. Recent work has also shown that the palm can support the perception of thermal cues \cite{filingeriThermosensoryMicromappingWarm2018,zhangThermalPerceptionInformation2022}.

We make use of the large surface area of the palm by incorporating an array of $6.5~\mathrm{mm} \times 6.5~\mathrm{mm}$ TEMs (CUI Device CP076581-238P) arranged in a $3 \times 3$ grid with $18~\mathrm{mm}$ spacing. Multi-actuator thermal arrays have been used in prior thermal rendering systems. Zhipeng et al. \cite{zhipengThermalTactileDisplay2012} developed a thermal display composed of a $3 \times 3$ set of heating actuators and a single TEM for cooling. More recently, Li et al. developed a wearable $4 \times 4$ array of thermoelectric actuators for hand rehabilitation after stroke \cite{liMuscleTemperatureSensing2020}. Kim et al. \cite{kimTwoDimensionalThermalHaptic2020} integrated a two-dimensional array of TEMs into the handle of a blind-assistive cane. \hl{These systems demonstrate the feasibility of array-based thermal rendering, but their use has primarily focused on non-AR or task-specific contexts rather than AR interactions that require simultaneous physical touch and virtual thermal feedback.}

The individual modules are encapsulated within a flexible silicone base, which was selected for mechanical flexibility and biocompatibility. The flexibility and lightness of the material allow the device to be worn against the palm while limiting interference with fingertip contact. A two-part mold fabrication process was used to create internal water-cooling channels that disperse heat away from the TEMs and help maintain actuator efficiency during sustained operation. The current prototype remains tethered because of the external control circuitry and water-cooling reservoir. When strapped to the hand, the flexible base conforms to the palm surface, improving contact consistency between the thermal array and the skin and reducing positional slippage during movement.

\hl{Prototype characterization showed that the device's operational temperature range spans approximately $15\celsius$ above and below the user's ambient skin temperature, with onset temperature change rates exceeding $0.1~\celsius/\mathrm{s}$ within $50~\mathrm{ms}$. The average delay from external-facing contact detection to actuation of the inward-facing TEMs against the palm was $5~\mathrm{ms}$. Combining this contact-detection delay with the actuator response delay resulted in an estimated physical surface rendering delay of $55~\mathrm{ms}$. This delay is below typical thermal sensing reaction times of the skin, which are on the order of $300$--$700~\mathrm{ms}$ \cite{yarnitskyWARMCOLDSPECIFIC1991,kenshaloWarmCoolThresholds1968}. A more detailed description of the device fabrication and performance characterization is provided in Section \ref{sec:Methods}: Methods.}

\FloatBarrier

\subsection{Participants}\label{subsec:Participants}
Twelve (12) individuals from Vanderbilt University's undergraduate and graduate student populations were recruited for this study with approval from Vanderbilt's Institutional Review Board. Recruitment focused on participants who were available during the study's scheduled sessions and willing to take part. Prior to participation, each individual was provided with a detailed explanation of the study's purpose, procedures, potential risks, benefits, and their rights as participants. Written informed consent was obtained after participants had the opportunity to ask questions and receive clarification.

Each participant attended four sessions, lasting one hour each, during which they completed the study's tasks and activities. This session length was chosen to balance data collection needs with minimizing participant fatigue or discomfort.

The participant group had a median age of 24.2 years with a standard deviation of 3.4 years. Most participants reported minimal or no prior experience with head-mounted display (HMD) VR/AR technology.

\subsection{Experiment 1: JND Testing} \label{ch4:JNDTesting}
\begin{figure}[htbp]
    \centering
    \includegraphics[width=0.56\textwidth]{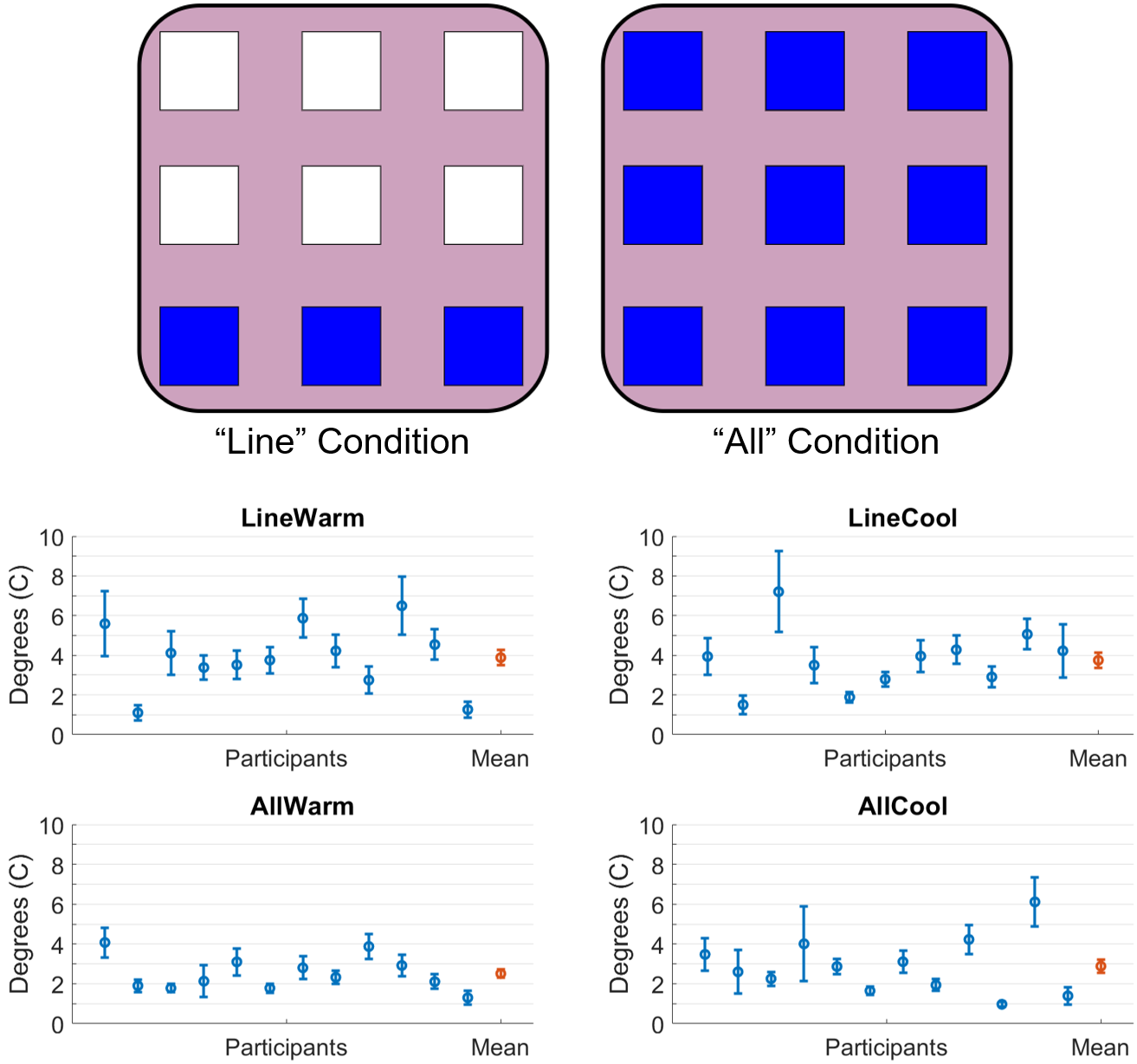}
    \caption{\textbf{Just Noticeable Difference (JND) Experiment Conditions and Results}\\ Two distinct spatial patterns were applied to the device for both heating and cooling JND calculations. (\emph{Line}) The "Line" pattern activated bottom row of TEMs on the device. (\emph{All}) For the "All" pattern, all TEMs were activated. (\emph{JND Results}) The JND (in degrees celsius) quantifies the smallest change in a reference temperature that is detectable with a 75\% confidence. Per-participant JND is shown in blue and the overall JND is shown in orange.}
    \label{fig:Ch4-FirstExperiment}
\end{figure}

\hl{To characterize whether thermal changes rendered through the device were perceptible, we quantified the just noticeable difference (JND) for device-mediated thermal stimuli. The JND represents the smallest difference between a reference stimulus and a test stimulus that can be detected with a specified level of reliability. In this experiment, we defined the JND as the stimulus difference corresponding to a 75\% probability that a participant would report the reference and test stimuli as different.}

\hl{We used a one-up/one-down adaptive weighted staircase procedure with an asymmetric step-size to target the 75\% detection threshold\cite{kingdomAdaptiveMethods}. On each trial, participants were presented with a reference stimulus followed by a test stimulus. The test stimulus differed from the reference stimulus by an adjustable percentage offset. After each response, this offset was updated based on the participant's judgment. If the participant reported that the stimuli were different, the offset was decreased by 10\%, making the next comparison more difficult. If the participant reported that the stimuli felt the same, the offset was increased by 30\%, making the next comparison easier.}

The adaptive staircase experiment was conducted for four total conditions: warming the entire thermal device, cooling the entire thermal device, warming a single row of the TEM array, and cooling a single row of the array. The bottom horizontal row of TEM elements in the device was chosen for the line conditions (Figure \ref{fig:Ch4-FirstExperiment} (a-b)), which corresponds to testing the JND of the thenar eminence and base of the palm. The chosen conditions were selected to evaluate both the global and localized thermal sensitivity of the palm, as these areas are critical for perceiving thermal feedback in AR applications.

A reference stimulus of $4\celsius$ above or below the device's ambient temperature ($30\celsius$) was chosen for the heating and cooling trials, respectively. An initial step size of $4\celsius$ was chosen, meaning the initial test stimulus was $8\celsius$ above or below ambient conditions. Each stimuli was presented to the user for a period of 3.5 seconds, with no Inter-Stimulus Interval (ISI) between them (ISI=0s). No ISI was chosen to avoid time order errors, based on the results of Hojatmadani et al. \cite{hojatmadaniTimeDelayAffects2022}, who found that ISI had a significant effect on calculated JNDs. Before and after the presentation of the two stimuli the thermal device was allowed to return to its ambient temperature. Each staircase test was conducted until 10 reversals were observed (changes in response from same to different or different to same, see Figure \ref{fig:Ch4-FirstExperiment} (c)). The step sizes at the last eight reversals were then averaged to calculate the participant's JND for the specific condition.

The results of the JND testing are shown in Figure \ref{fig:Ch4-FirstExperiment}. The individual JNDs for each condition failed to reject the null hypothesis of the Shapiro-Wilk test, indicating that they come from normal distributions. As such, the individual JNDs for each condition were fit to a normal curve in order to derive the 95\% Confidence Interval (CI). The "Line-Warming" condition was found to have an average JND of $3.88\celsius$ (95\% CI: $3.49\celsius$-$4.27\celsius$), the "Line-Cooling" condition to have an average JND of $3.75\celsius$ (95\% CI: $3.36\celsius$-$4.14\celsius$), the "All-Warm" condition to have an average JND of $2.51\celsius$ (95\% CI: $2.30\celsius$-$2.71\celsius$), and the "All-Cool" condition an average JND of $2.88\celsius$ (95\% CI: $2.54\celsius$-$3.22\celsius$). This is consistent with previous work by Hojatmadani et al. \cite{hojatmadaniTimeDelayAffects2022} investigating the JND of the human palm.

An ANOVA was run on all four conditions and no significant difference was found among the conditions ($p=0.0549$). However, the low p value gives weight to the observed trend that the JND was smaller for the "All" mode conditions, implying that users can detect a smaller threshold between presented stimuli when the entire TEM array is activated.

\FloatBarrier
\newpage

\subsection{\hl{Experiment 2: User Experience with Indirect Feedback}}

\begin{figure}[htbp]
    \centering
    \includegraphics[width=\textwidth]{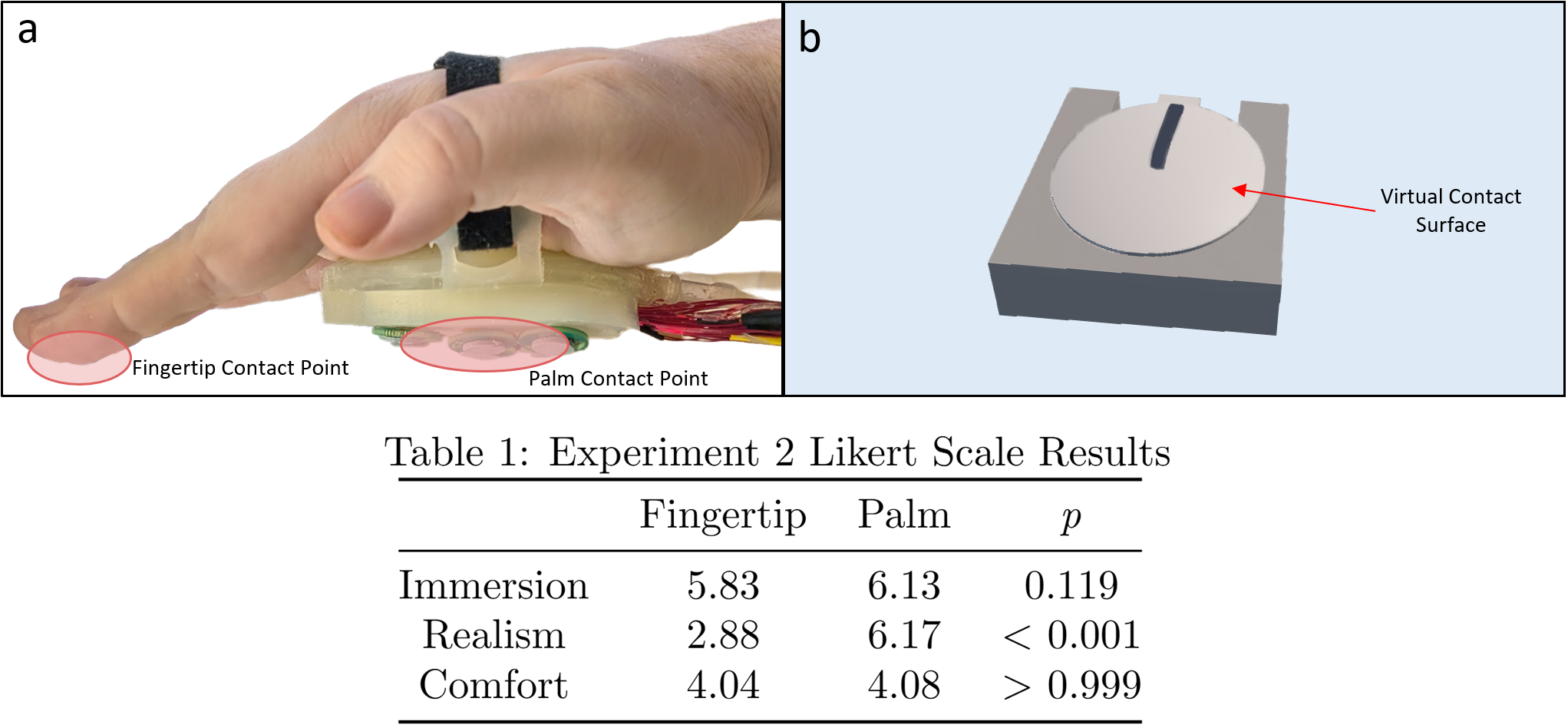}
    \caption{\textbf{Indirect Feedback Experimental Setup and Results.}\\ \hl{(\emph{a}) The fingertip and palm contact points utilized in this experiment. In both scenarios thermal sensations are applied via our device to the palm contact region. (\emph{b}) The virtual object with an identifiable contact surface utilized for this experiment. Participants touched the central circular region with both their fingertips and palm. (\emph{Table 1}) Likert-scale results measuring immersion, realism, and comfort. The observed mismatch between the fingertip and palm conditions in perceived realism, i.e. how accurate the sensations were compared to a real interaction, did not affect the comfort of the interaction or its immersion, i.e. suspending disbelief to perceive the virtual object as actually being the sensed temperature.}}
    \label{fig:Exp2}
\end{figure}
\FloatBarrier

\hl{Although indirect feedback can preserve dexterity by leaving the fingertips unobstructed, this design choice requires that thermal cues delivered to the palm remain immersive during interactions that visually occur at the fingertips. It also requires that the mismatch between visual contact location and sensory feedback location does not increase user discomfort or decrease immersion . This experiment aims to validate \textbf{DC 1: Indirect Feedback} by evaluating whether thermal feedback applied to the palm can support fingertip-based virtual interactions without reducing the perceived quality of the experience.}

\hl{Participants wore our haptic device and interacted with a virtual object set to either a warm or cool surface temperature. Each participant completed four trials: two fingertip interaction trials and two palm interaction trials. For each interaction location, one trial presented a warm stimulus and one presented a cool stimulus. The temperature setpoint for each trial was defined relative to the participant's skin temperature, with warm trials set to $8\celsius$ above skin temperature and cool trials set to $8\celsius$ below skin temperature. Trial order was randomized within participants.}

\hl{After completing all four interaction trials, participants answered three Likert-scale questions for both the fingertip and palm interaction conditions. Participants rated their sense of immersion, the realism of the thermal sensations, and their comfort during the interaction. Each question used a 1--7 scale, where 1 indicated no immersion, no realism, or maximum discomfort, and 7 indicated complete immersion, complete realism, or complete comfort.}

\hl{Table 1 in Figure \ref{fig:Exp2} shows the results of the Likert-scale responses. Wilcoxon Signed Rank tests were conducted to compare responses between the fingertip and palm interaction conditions. No significant difference in immersion was observed ($p=0.119$), with an average rating of 5.83 for fingertip interactions and 6.13 for palm interactions. Comfort ratings additionally showed no significant difference ($p>0.999$), with an average score of 4.04 for fingertip interactions and 4.08 for palm interactions.}

\hl{In contrast, realism ratings were significantly lower for fingertip interactions than for palm interactions ($p<0.001$). Fingertip interactions received an average realism score of 2.88, while palm interactions received an average realism score of 6.17. These results indicate that participants perceived a clear reduction in realism when thermal feedback was delivered to the palm while visual contact occurred at the fingertips. However, this mismatch did not transfer to reduced immersion or increased discomfort. Participants remained immersed despite the reduced realism of the indirect feedback condition, and comfort ratings were neutral in both conditions.}

\FloatBarrier

\subsection{\hl{Experiment 3: Active Temperature Passthrough}} \label{ch4:Passthrough}

\begin{figure}[htbp]
    \centering
    \includegraphics[width=\textwidth]{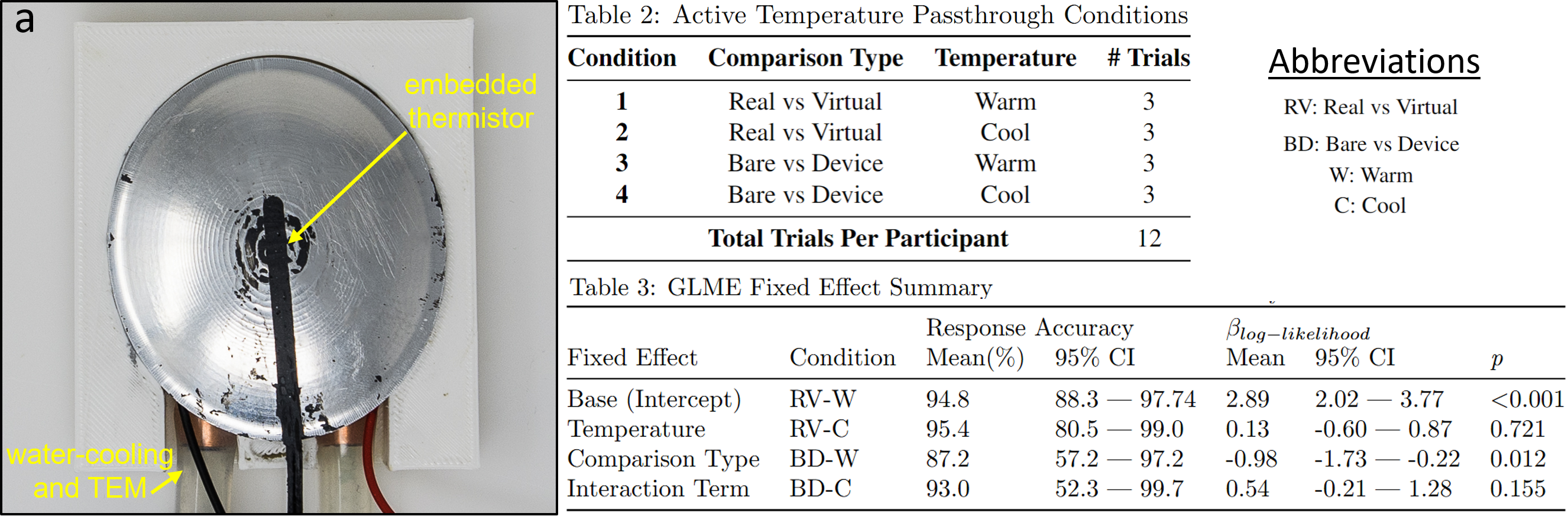}
    \caption{\textbf{Temperature Passthrough Experimental Setup, Conditions, and Results.}\\ (\emph{a}) A convex aluminum surface with an embedded thermistor sits atop a thermoelectric module and heatsink to act as a temperature-controlled surface. (\emph{Table 1}) Two conditions--the temperature of the surface relative to ambient and the method of comparison--were tested using a Generalized Linear Mixed Effects model (\emph{Table 2}).}
    \label{fig:Exp3}
\end{figure}
\FloatBarrier

\hl{No matter the choice of body placement of an AR indirect thermal feedback device, there will be an occluded area of skin the user cannot use for sensing their physical environment. This experiment validates \textbf{DC 2: Temperature Passthrough} by characterizing the efficacy and immersiveness of our thermal device's "temperature pass-through" capability. We conducted an object interaction experiment utilizing a test stand with a temperature-controlled surface and virtual replicas of the test stand. The test stand's contact area was made from 6061 aluminum CNC'd into a concave ellipsoidal surface. An embedded thermistor in the aluminum enabled closed-loop temperature control using a 45mmx45mm TEM (ATS-TEC40-39-004) under the aluminum component. This assembly was then mounted onto a copper water-cooling heatsink for temperature management.}

\hl{Participants were asked to compare the temperature of a pair of objects and state whether the objects were at equal or different temperatures. Table 2 in Figure \ref{fig:Exp3} presents the tested conditions. The "Real vs Virtual" trials asked participants to wear the device on their right hand and compare the test stand's temperature against a virtual replica of the test stand. The "Bare vs Device" trials involved just the physical test stand; participants first touched the test stand with their bare left palm and then again with the haptic device equipped on their right palm. The bare left palm was used to avoid the long delays that would have been required to repeatedly mount and dismount the thermal device on the right hand.}

\hl{Each of the four conditions shown in Table 1 included three trials. In one trial, both object interactions were set to the same temperature offset of $5\celsius$ from the participant's skin temperature: a positive offset for Warm conditions and a negative offset for Cool conditions. For another trial one object interaction had an offset of $5\celsius$ and the other $10\celsius$, and in a third trial, object interactions had offsets of $5\celsius$ and $15\celsius$.}

\hl{The order in which each condition was tested was randomized between participants, and the order of the three temperature offset configurations was randomized within each condition. A 3 second delay was enforced between each trial's two object interactions so that the real test stand had sufficient time to change its surface temperature during the "Bare vs Device" condition. At the end of this experiment, participants were asked to rate the overall realism of the object interactions during the AR experience using a Likert-scale of 1-7, with 1 being completely unrealistic and 7 being completely realistic.}

\hl{To account for participant-level variation and within-subject temperature offset variation, a Generalized Linear Mixed Effects (GLME) model was used to predict response accuracy. Fixed effects included the warm vs cool condition, and the type of comparison being made ("Real vs Virtual" or "Bare vs Device"). Random effects were participant and temperature offset. A logit link function was employed as both predictors were ordinal. Significance of the Likert-scale realism score was tested using the Wilcoxon Sign Rank test. Post-hoc binomial tests were conducted to calculate response accuracy p-values.}

\hl{Table 3 in Figure \ref{fig:Exp3} shows the results of fitting the GLME model. Fixed effects showed that the warm vs cool condition had no significant effect on accuracy ($p=0.721$), while the bare-handed vs purely haptic predictor was significantly associated with accuracy ($\beta_{log-likelihood}=-0.976, p=0.012$). This effect can be seen in its average response accuracies, with the overall accuracy of the haptic device at $95.0\%$ ($p<0.001$), and the bare-handed vs device accuracy of $90.1\%$ ($p<0.001$). No significant interaction effect was observed. Realism scores had an average value of $5.75$ ($p<0.001$), indicating that participants experienced a moderate degree of realism.}

\FloatBarrier

\subsection{\hl{Experiment 4: Spatial Patterns}} \label{ch4:SpatialIcons}

\begin{figure}[htbp]
\centering
\includegraphics[width=\textwidth]{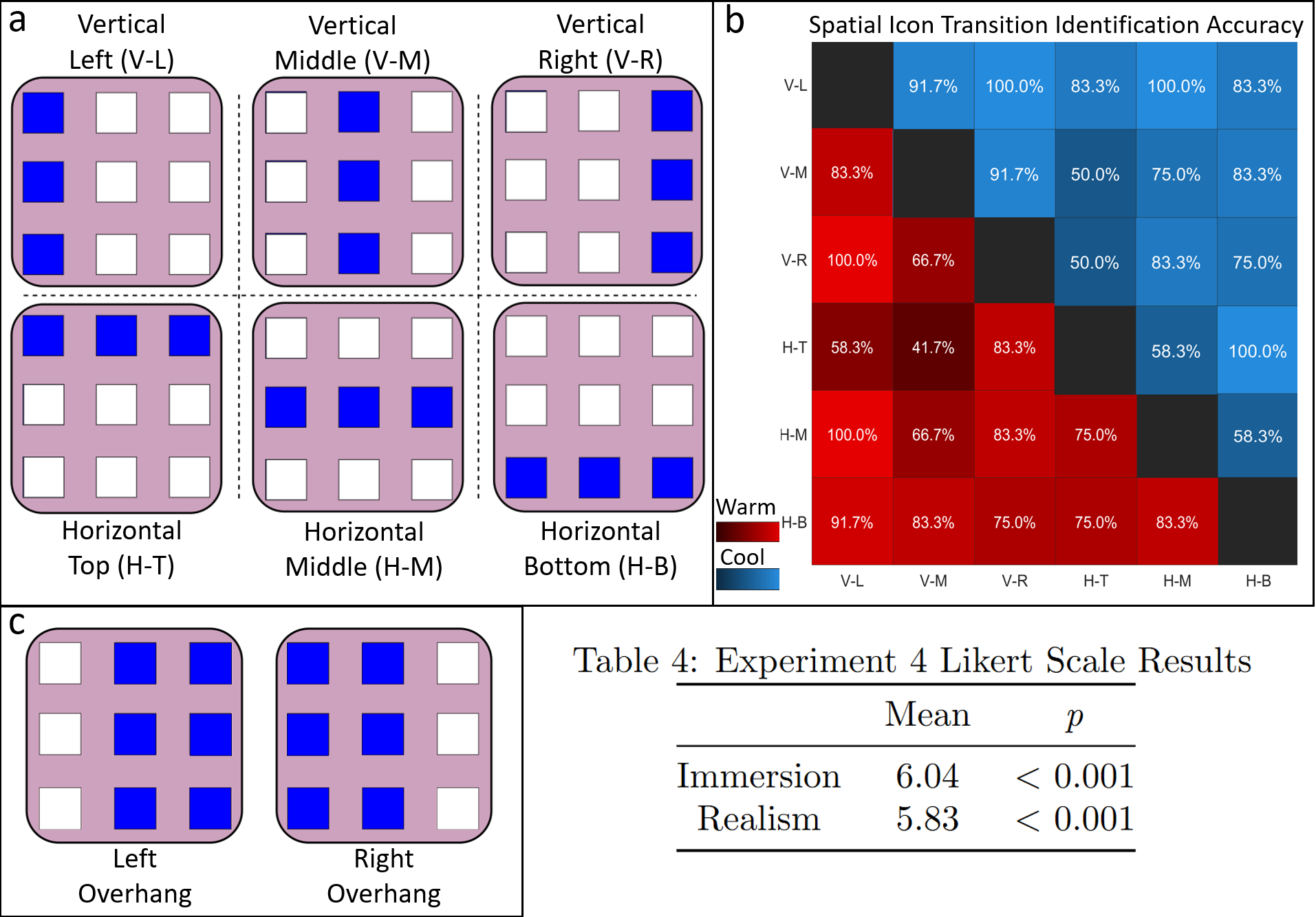}
\caption{\textbf{Spatial Patterns Experimental Setup and Results.}\\ \emph{(a)} Transitions between six spatial patterns that utilized the device's 3x3 TEM grid were applied to participants wearing the device, for both warm and cool sensations. \emph{(b)} The reported discrimination accuracy for each pattern pair transition are shown in the matrix heatmap. \hl{\emph{(c)} The thermal patterns corresponding to the participant's palm overhanging the left or right side of the virtual object. \emph{(Table 4)} Likert-Scale results for Immersion and Realism.}}
\label{fig:Exp4}
\end{figure}
\FloatBarrier

\hl{To begin validating \textbf{DC 3: Spatiotemporal Patterns}, we first evaluated participants' spatial discrimination accuracy. We developed a spatial pattern discrimination task in which participants were presented with one of six predefined patterns: the left, right, or middle vertical column of the array, or the top, bottom, or middle horizontal row of the array. After a brief period of exposure, the pattern either transitioned to another pattern from the set or continued to present the same thermal pattern. Participants were asked to report whether they noticed any change in the spatial profile of the thermal sensation applied to their palm. This process was conducted for every combination of pattern pairs from the set, for both warm and cool stimuli.}

\hl{A temperature step size of $8\celsius$ above or below skin temperature was used for the presentation of thermal patterns. This was chosen to be well above the JNDs measured in the first experiment to ensure that discrimination was not limited by insufficient temperature differences. Each pattern was applied to the participant for 3 seconds before either transitioning to the second pattern for 3 seconds or maintaining the current pattern for an additional 3 seconds. The thermal device then returned to ambient skin conditions. A total of 60 trials were conducted with each participant: 15 trials with a spatial pattern transition and 15 trials without a transition, for both the warm and cool conditions.}

\hl{The results of the spatial pattern discrimination task are shown in Figure \ref{fig:Exp4}. Accuracy in detecting pattern changes ranged from 41.7\% to 100\% for warm stimuli and from 50.0\% to 100\% for cool stimuli. The warm condition had an overall transition detection accuracy of 77.8\%, while the cool condition had an overall transition detection accuracy of 78.9\%. There was no significant difference in responses between the warm and cool conditions ($p>0.999$).}

\hl{We also evaluated whether spatially rendered thermal feedback could support immersive virtual object interactions. Participants were asked to press their palm against a virtual replica of the temperature-controlled test stand used in Experiment 3, with the hand positioned such that approximately half of the palm contacted the virtual object and the remaining half extended beyond either the left or right side of the object. When combined with hand tracking, this interaction activated either one or two rows of the thermal array depending on variation in hand placement. Each participant experienced this interaction twice: once with a warm stimulus set to $8\celsius$ above skin temperature and once with a cool stimulus set to $8\celsius$ below skin temperature. The order of warm and cool stimuli and left- and right-sided placement of the palm were randomized within participants.}

\hl{After completing both interactions, participants answered the same Likert-scale questions on immersion and realism used in Experiment 2, show in Table 4 in Figure \ref{fig:Exp4}. Single-sample Wilcoxon Signed Rank tests were used to evaluate whether the ratings were significantly different than the neutral midpoint of the scale. Participants reported an average immersion score of 6.04 ($p<0.001$) and an average realism score of 5.83 ($p<0.001$), indicating that spatially rendered thermal feedback supported both immersive and realistic object interactions. These ratings were not significantly different from the palm interaction ratings in Experiment 2 for either immersion ($p=0.617$) or realism ($p=0.135$) (Table 1 in Figure \ref{fig:Exp2}, suggesting that spatially constrained thermal rendering preserved the perceived quality of direct palm-based feedback.}

\FloatBarrier

\subsection{\hl{Experiment 5: Moving Thermal Rendering in AR}} \label{ch4:MovingThermalRendering}

\begin{figure}[htbp]
\centering
\includegraphics[width=\textwidth]{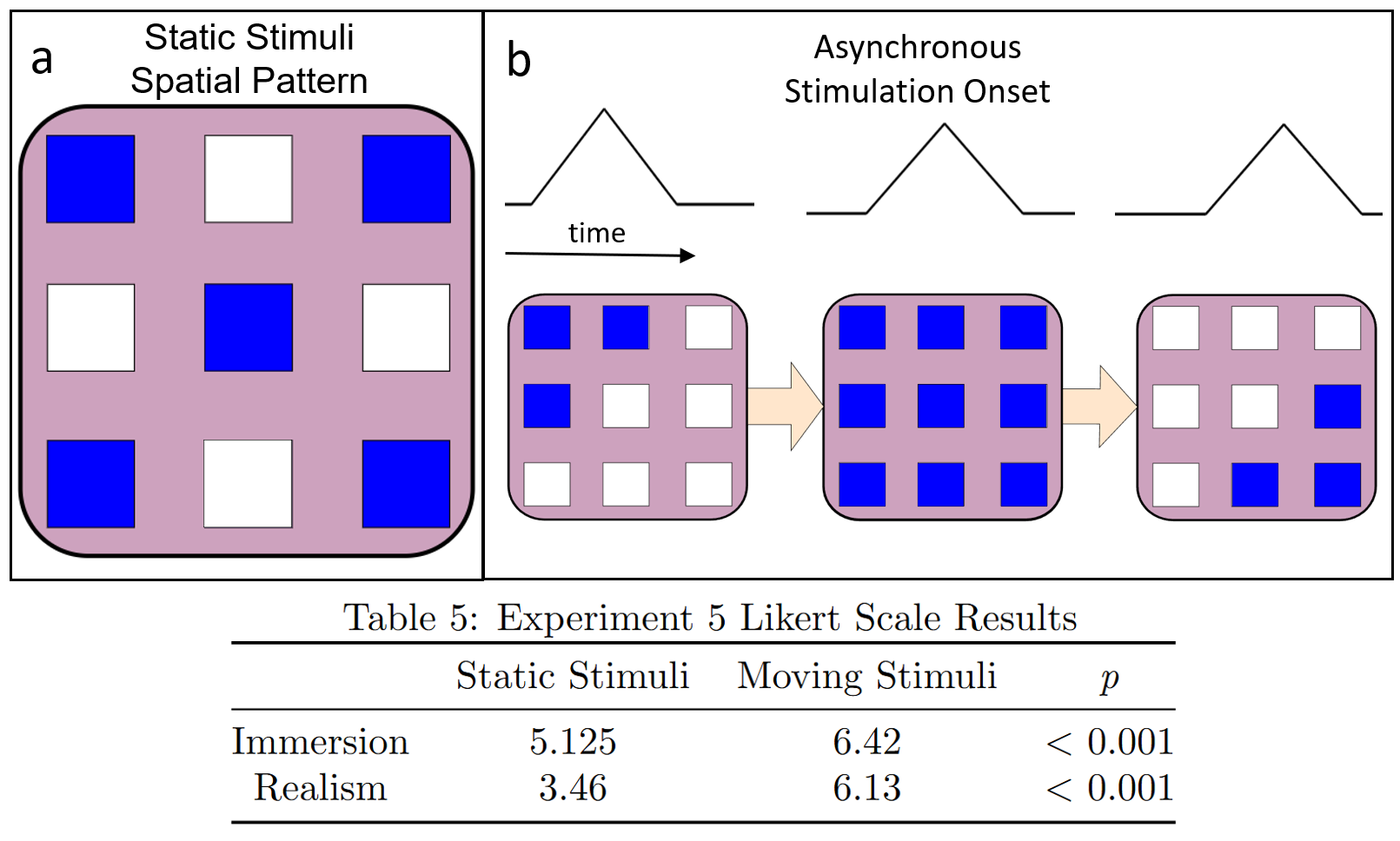}
\caption{\hl{\textbf{Moving Stimuli Experimental Setup and Results.}}\\ \hl{\emph{(a)} A static cross-shaped pattern was used at the control condition for this experiment. \emph{(b)} An example depiction of a moving spatiotemporal pattern created by the asynchronous activation of TEMS as the user's hand and haptic device move laterally across a virtual surface. \emph{(Table 4)} Likert-Scale results for Immersion and Realism.}}
\label{fig:Exp5}
\end{figure}
\FloatBarrier

\hl{To finish validation of \textbf{DC 3: Spatiotemporal Patterns}, a final experiment was conducted to evaluate immersion and realism during a dynamic spatiotemporal AR object interaction (Figure \ref{fig:Exp5}). While still wearing the device, participants were asked to place their right palm to the left of --- and level with the surface of --- a virtual replica used in previous experiments of the temperature-controlled test stand used in Experiment 3. Participants were then asked to brush their palm laterally from left to right across the surface of the virtual object until their palm was positioned to the right of the object. They were instructed to perform this brushing motion at a slow pace, with approximately 3 seconds elapsing from initial contact on the left side of the virtual surface to the end of contact on the right side. The motion and speed were demonstrated by study personnel for each participant before they were asked to perform the action.}

\hl{Two thermal rendering conditions (Figure \ref{fig:Exp5} a,b) were tested: a static thermal stimulus and a moving thermal stimulus. In the static condition, thermal feedback was applied to the palm by activating the TEMs at the center and four corners of our haptic device's array when the participant's palm made contact with the virtual object's surface and deactivating the TEMs when contact stopped. In the moving condition, TEM actuators in the array were dynamically activated based on hand tracking and the estimated overlap between the participant's palm and the virtual object, creating a spatially varying thermal pattern as the hand moved across the surface. Participants were able to experience each condition once with a warm stimulus set to $8\celsius$ above skin temperature and once with a cool stimulus set to $8\celsius$ below skin temperature. The order of warm and cool stimuli were randomized within participants.}

\hl{After completing the interaction trials, participants answered Likert-scale questions on immersion and realism for both the static and moving thermal rendering conditions. The Wilcoxon Signed Rank test was again utilized for analysis. The moving thermal rendering condition produced significantly higher ratings than the static condition for both immersion ($p<0.001$) and realism ($p<0.001$). Average immersion increased from 5.13 in the static condition to 6.42 in the moving condition, while average realism increased from 3.46 in the static condition to 6.13 in the moving condition.}

\hl{These results indicate that spatiotemporal thermal rendering improved the perceived quality of dynamic AR object interactions. Compared with a static stimulus applied to the center of the palm, thermal feedback that moved across the array in correspondence with hand-object overlap produced higher immersion and realism, supporting the use of spatially and temporally varying thermal patterns for rendering contact with virtual surfaces.}

\FloatBarrier
\section{\hl{Discussion}} \label{sec:Discussion}

\hl{We developed and evaluated a palm-mounted thermal feedback prototype for augmented reality that integrates active heating and cooling, real-time temperature sensing, and independently controlled spatial thermal rendering. The device was designed around three AR-specific design considerations: preserving manual dexterity through indirect feedback, maintaining access to real-world temperature information through thermal passthrough, and rendering spatially and temporally dynamic thermal cues for virtual object interactions. Across five controlled human-subject experiments, the results provide evidence that palm-mounted thermal feedback can support immersive AR interactions while preserving important aspects of real-world touch.}

\subsection{\hl{Indirect feedback and the tradeoff between dexterity and realism}}

\hl{Experiment 2 demonstrates an important distinction between immersion and realism in indirect thermal feedback. Delivering thermal feedback to the palm while visual contact occurred at the fingertips significantly reduced perceived realism, indicating that participants recognized the mismatch between where the interaction appeared to occur and where the thermal sensation was felt. However, this mismatch did not significantly reduce immersion or increase discomfort. This suggests that exact colocation of thermal feedback may not be necessary for every AR interaction.}

\hl{This finding has practical implications for wearable thermal device design. Fingertip-mounted devices can provide spatially congruent feedback, but they also interfere with the same contact surfaces that users rely on for grasping, manipulation, and texture exploration. Palm-based indirect feedback can be a valid design alternative: it preserves the fingertips for natural interaction while still providing thermal cues that can support the user's sense of engagement with the virtual environment. The reduced realism observed in the fingertip condition indicates that this approach is not equivalent to direct fingertip feedback. However, the preserved immersion suggests that users may tolerate this sensory displacement when the goal is to enhance the general sense of immersion in AR experience rather than perfectly reproduce fingertip contact. This distinction is also particularly relevant for AR applications where users frequently transition between real and virtual interactions. In these scenarios, preserving manual dexterity may be more important than achieving perfect sensory colocation. Conversely, applications requiring precise fingertip realism, such as fine material discrimination or high-fidelity manual training, may require either fingertip feedback or a hybrid approach that combines direct and indirect stimulation. Together, these results support \textbf{DC 1: Indirect Feedback} by showing that thermal feedback can be relocated away from the apparent point of contact without reducing immersion or increasing discomfort. }

\subsection{\hl{Active Thermal passthrough and preserving real-world interaction}}

\hl{The passthrough results support the value of treating occlusion as a primary design problem for AR haptics. Any wearable device placed on the hand blocks some portion of natural tactile and thermal perception. For a palm-mounted thermal device, this is especially important because the palm is often used for sustained contact with objects and surfaces. Without passthrough, the device would improve virtual interaction while degrading physical interaction, undermining the central goal of AR: blending virtual content with the real world.}

\hl{Experiment 3 indicates that temperature passthrough can restore a substantial portion of the thermal information lost when the palm is covered by the device. Participants were able to distinguish temperature differences with high accuracy when comparing physical and rendered thermal stimuli, and the reported realism scores suggest that the rendered sensations were perceptually convincing during object interactions. The lower performance in the Bare vs Device comparison, relative to device-mediated comparisons, also shows that passthrough does not fully reproduce bare-skin sensing. This is expected because the device inserts several stages between the object and the user: external temperature sensing, control logic, actuator response, heat transfer through the TEM interface, and contact with the skin.}

\hl{The implication is not that passthrough eliminates the perceptual cost of wearing a device, but that it can reduce this cost enough to preserve usable thermal interaction with the physical environment. This is an important design direction for AR haptics. Rather than choosing between augmenting virtual objects and preserving real-world touch, future devices can actively mediate between the two. }

\hl{These results support \textbf{DC 2: Temperature Passthrough} by showing that thermal passthrough can preserve meaningful access to real-world temperature cues even when the skin is physically occluded by a wearable device. While passthrough does not fully reproduce bare-skin perception, it provides a practical mechanism for maintaining thermal interaction with physical objects while still allowing the device to render augmented or virtual thermal feedback.}

\subsection{\hl{Spatial and spatiotemporal rendering}}

\hl{Experiment 4 showed that participants were able to detect transitions between spatial thermal patterns at rates consistent with previous studies involving stationary thermal arrays \cite{zhangThermalPerceptionInformation2022,yangQuantitativeTactileDisplay2006}. This suggests that even a low-resolution thermal array can display spatial cues that extend beyond simply rendering the bulk thermal properties of an object. The ability to discriminate between thermal patterns supports the use of selectively activated regions of the TEM array to represent partial or irregular contact with a virtual object. For example, if a user touched a virtual cup partially filled with hot liquid, only the portion of the palm aligned with the filled region would need to be stimulated. The high immersion and realism ratings observed during the spatial rendering task further suggest that this type of spatially constrained feedback can preserve the perceived quality of palm-based thermal interaction.}

\hl{Experiment 5 extended this concept to dynamic contact. When participants brushed their palm across a virtual surface, moving thermal feedback based on hand-object overlap produced significantly higher immersion and realism than a static stimulus applied to the center of the palm. This supports the idea that spatiotemporal thermal patterns can improve AR interactions by making the rendered feedback correspond to both contact location and hand motion. Rather than only indicating that a virtual object is warm or cool, the device can update the location of thermal stimulation as the interaction unfolds.}

\hl{Together, these results support \textbf{DC 3: Spatiotemporal Patterns} by showing that spatially and temporally varying thermal cues can contribute to more immersive and realistic AR object interactions. They also suggest that dynamic thermal effects can be produced using thermal actuation alone, without requiring an additional vibrotactile modality to imply motion.}

\subsection{\hl{Design implications for AR thermal haptics}}

\hl{A central implication of these findings is that thermal feedback in AR should not be treated only as a question of actuator performance or maximum temperature range. Instead, the effectiveness of a thermal haptic system depends on how well the device fits within the user's broader interaction context. In AR, the user must continue to manipulate physical objects, sense real-world surfaces, and maintain perceptual coherence between physical and virtual elements. The results of this work suggest that these requirements can be addressed through a combined approach: relocating thermal feedback away from dexterous contact regions, restoring occluded thermal information through passthrough sensing, and dynamically rendering thermal patterns that correspond to virtual contact geometry and motion.}

\hl{While indirect feedback is viable, its role should be chosen carefully. It can preserve immersion and comfort while maintaining manual dexterity, but it may reduce realism when the apparent contact point is far from the stimulated skin region. Designers should therefore use indirect thermal feedback when the benefit of leaving the fingers or other critical body part unobstructed outweighs the need for exact sensory colocation.}

\hl{For AR applications, passthrough should be considered a core requirement for wearable thermal devices that cover thermally sensitive areas of skin. A device that only renders virtual temperature cues risks degrading the user's ability to perceive the real environment. By sensing and reproducing contacted surface temperatures, passthrough allows the device to serve both as an augmentation tool and as a compensatory interface for the skin it occludes.}

\hl{Finally, spatial and temporal rendering should be prioritized not constrained by device size or design complexity. Strong, static thermal stimuli may be easier to produce and control, but they do not necessarily improve coherence with AR interactions. The results from Experiments 4 and 5 suggest that thermal feedback becomes more compelling when it corresponds to contact location, object geometry, and hand motion. This points toward rendering strategies that are contact-aware rather than stimulus-only: the device should not merely output a target temperature, but should determine where, when, and how that temperature should appear on the hand.}

\subsection{\hl{Limitations and Future Work}}
\hl{Although wearable, the current prototype remains tethered to bench-top control electronics and a water reservoir, constraining user mobility to a tabletop workspace. Future iterations should integrate temperature management solutions onto the forearm, potentially via phase-change cooling methods \cite{jungHighHeatStoring2023}, to aid portability. If water-cooling or other fluid-based cooling continues to be necessary, a possible space-saving technique could be the removal of a bulky fluid reservoir and instead relying on heat-exchange between fluid in the system's tubing and the skin of the user's forearm. Combining such as human skin heat-exchange method with new advances in wearable technology for human thermo-regulation \cite{luoFlexibleLiquidMetalbased2024} could avoid issues with localized temperature discomfort. Full portability could then be achieved by relocated Heavy and bulky items such as batteries and control circuitry to a harness, belt, or backpack attached to the user's torso, with minimal cabling extending to the primary device on the palm and forearm.}

\hl{Additionally, uneven contact pressure across the thermoelectric array produced variability in response times. Attempts were made to allow the TEM array to contour to the palm by using a flexible silicone material and molding it to have a convex shape. Yet it is apparent by looking at the system response of individual TEMS (Figure \ref{fig:Ch4-TDFLIR-SystemResponse} (d) and (e)) that not all TEMs exhibited the same thermal response times and damping/underdamping behavior when in contact with the palm. Contact pressure has long been known to affect thermal resistance between an object and human skin \cite{chenFingerCoolingContact1994,rykaczewskiModelingThermalContact2019,tiestExperimentallyVerifiedModel2007}. A variable contact pressure then affects the heat flow out of a TEM and into the surrounding skin. This means that a set of PID gains for temperature control for one set of TEMs will not be appropriate for another set with a different average contact resistance. A more efficient way of applying even contact pressure is needed, or a method of determining thermal contact resistance and adjusting individual PID gains accordingly.}

\hl{In conclusion, this work provides evidence supporting the three proposed design considerations for thermal feedback in AR. Indirect feedback can preserve manual dexterity while maintaining immersion, thermal passthrough can reduce the perceptual cost of occluding the skin, and spatiotemporal rendering can improve the realism of virtual object interactions by aligning thermal cues with contact location and motion. Taken together, these findings suggest that effective AR thermal haptics require more than the addition of temperature cues; they require feedback strategies that intentionally balance physical interaction, real-world thermal perception, and coherent virtual sensation.}
\section{Methods} \label{sec:Methods}

\subsection{IRB Approval}\label{subsec:IRB}

This research study underwent ethical review and was approved by the appropriate Institutional Review Board (IRB) prior to its initiation. The purpose of the IRB review is to ensure the protection of human participants involved in the research and to ensure compliance with applicable ethical guidelines and regulations.

\subsection{AR Hardware and Development Software}\label{subsec:ARTools}
All AR testing was performed with the HoloLens 2 headset. The AR scenarios used in experiments 1 through 4 were developed using Unity 2020.3.34f1 and the Microsoft Mixed Reality Toolkit (MRTK) v2.8.3.

The thermal haptic device was fabricated using a three-layer silicone structure, each layer serving a specific function. The top layer, which made direct contact with the user’s palm, was designed to accommodate the TEMs and heat sinks. The middle layer incorporated water channels for cooling the cold side of the TEMs and provided attachment points for the Velcro strap used to secure the device to the user’s hand. The bottom layer provided space for wire management and included dedicated slots for the thermistors, which measured the temperature of the real-world objects with which the user interacted.

Each silicone layer was molded using two-part silicone and fabricated in a two-part 3D-printed mold. These layers were cast separately and bonded with silicone-compatible CA glue. The heatsinks were inserted into the slots in the top silicone layer and attached with CA glue, and the TEMs were attached to the heatsinks using thermally conductive epoxy. The same epoxy was also used to affix a thermistor on top of each TEM to provide control loop feedback. Thermistor slots were strategically positioned in the bottom layer directly below each TEMs to provide a one-to-one mapping of the TEMs with each external thermistor. This allowed the device to achieve "thermal passthrough" and pass on the temperature measured from the external object to the user's palm without loss of spatial resolution.

\subsection{Thermoelectric Modules} \label{ch4:TEMs}

\begin{figure}[htbp]
    \centering
    \includegraphics[width=\textwidth]{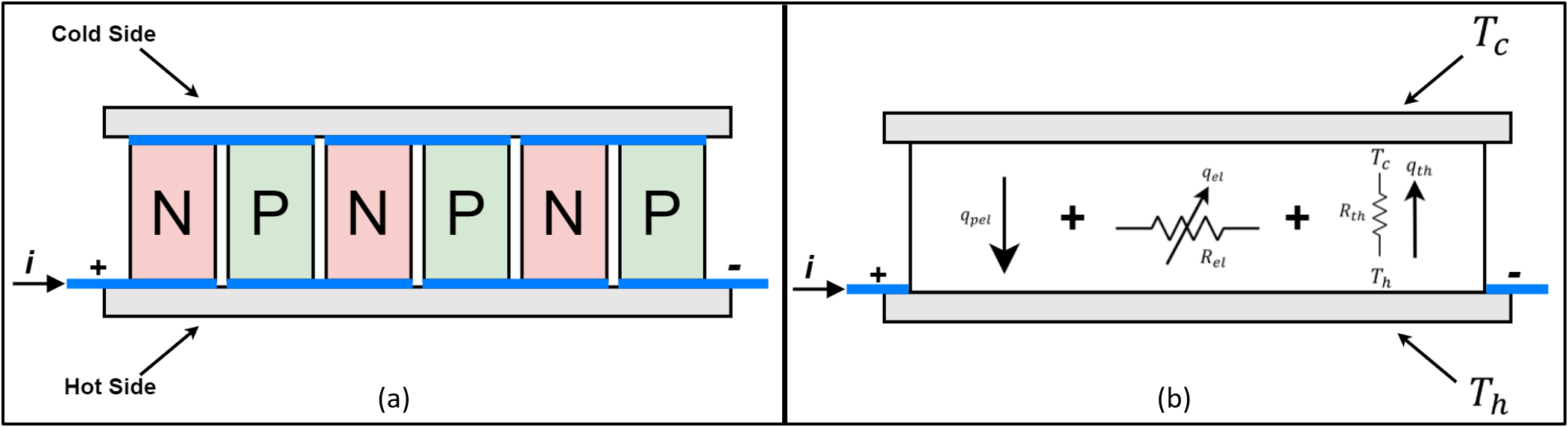}
    \caption{\textbf{Thermoelectric Module Overview.}\\ \emph{(a)} A thermoelectric device is made from alternating n- and
    p-type semiconductors placed in series electrically but in parallel
    thermally. When a DC current is passed through the semiconductors, a hot
    and cold side develop. \emph{(b)} This temperature differential is due to heat
    flow from the Peltier effect, \(q_{\text{pel}}\), and is resisted by
    electrical waste heat generation, \(q_{\text{el}}\), and natural heat
    conduction, \(q_{\text{th}}\).}
    \label{fig:Ch4-TEMOverview}
\end{figure}

A thermoelectric module, also known as a Peltier device, creates heat
flux across the junction of two dissimilar materials. When a DC electric
current flows through the device, a temperature differential arises
across the module as heat is transferred from one side to another,
creating a ``hot'' side and a ``cold'' side, shown in Figure \ref{fig:Ch4-TEMOverview} (a).
The direction of heat flow is reversible by changing direction of
current flow through the TEM, meaning that a single contact surface can
be used for both heating and cooling an object. This is important to keep
in mind as the terms ``hot'' and ``cold'' are used to reference specific
surfaces of the TEM, regardless of the direction of the temperature
gradient. Convention is to designate the temperature-controlled surface
as the ``cold'' side, with the ``hot'' side then being used for thermal
management.

Heat absorption on the ``cold'' side of the device (\(q_{\text{pel}})\)
is a function of component materials, current temperature (\(T_{c}\))
and applied current (\(I\)):

\begin{equation} \label{eq:ch4-peltiereq}    
  q_{\text{pel}} = \  - \alpha T_{c}I
\end{equation}

\noindent where \(\alpha\) is the Seebeck coefficient of the device, a term
that encapsulates material properties, and the negative sign represents
heat flow \emph{out} of the ``cold'' side. This heat absorption is
impeded by two additional effects: natural heat conduction back to the
``cold'' side due to the temperature differential, and waste heat
generated due to the device's internal resistance and applied current.

Actual heat flow out of the ``cold'' side is more accurately described
using the components of Figure \ref{fig:Ch4-TEMOverview}(b) in the following equation:

\begin{equation} \label{eq:ch4-temheatflow}
  q_{c} = \  - \alpha T_{c}I + \frac{1}{R_{\text{th}}}\left( T_{h} - T_{c} \right) + \frac{1}{2}R_{\text{el}}I^{2}
\end{equation}

where \(R_{\text{th}}\) and \(R_{\text{el}}\) represent the
device's thermal and electrical resistances, respectively. Only half of
the electrical heat generated,\(\ R_{\text{el}}I^{2}\), is
included, as it is assumed waste heat is dumped equally on both sides of
the module. The last two terms describing thermal conduction and waste
heat have an important ramification: as the temperature differential
across the device rises a decreasing amount of heat is effectively moved
and the module becomes less efficient. There comes a point where natural
heat conduction and waste heat generation completely negate the heat
flow induced by the thermoelectric effect and the ``cold'' side
temperature begins to rise, defining a maximum \(\Delta T\) that
can be generated across the device. If the ``hot'' side of the device is
held near ambient conditions using a heat sink or other thermal
management solution, this maximum temperature differential describes the
bounds of simulated thermal sensations that can be applied to the skin
contacting the device.

\subsection{Control Architecture} \label{ch4:ControlArchitecture}

Special consideration must be given to the control hardware used
with a TEM. Voltage regulation using an H-Bridge circuit controlled with
pulse-width modulation (PWM) has a negative impact on the TEM's
efficiency due to excessive waste heat generation. As described in equation
(\ref{eq:ch4-temheatflow}), this is because the waste heat generated scales
quadratically with applied current while thermoelectric heat flux is
only linearly proportional to current. A PWM controller works by rapidly
switching between a constant DC voltage and a grounded input. This
method works well for loads with an inductive component, such as a
motor, as the resulting current across the load is effectively steady
and scales with the PWM's duty cycle. A TEM, on the other hand, is a
purely resistive load, meaning that the induced current rapidly switches
between a max value and zero amps, amplifying the waste heat losses. The
solution to this problem is to filter the TEM's power source to generate
a variable DC voltage supply with as little ripple voltage as possible. This can be implemented using a differential filter or a class D amplifier, but is more easily achieved by utilizing off-the-shelf TEM drivers and additional passive components to complete the filtering circuit.

\begin{figure}[htbp]
    \centering
    \includegraphics[width=\textwidth]{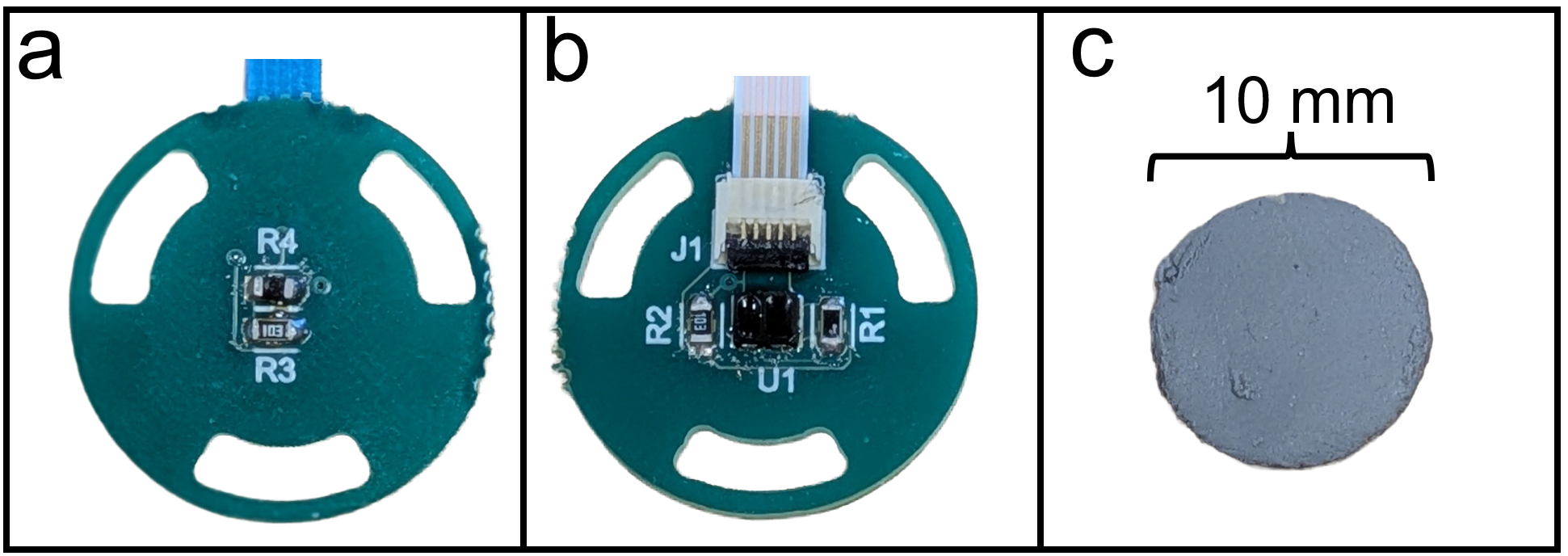}
    \caption{\hl{\textbf{Surface Contact and Temperature Measurement Sensors.}}\\ \hl{(\emph{a}) An 0603 footprint 10k thermistor (R4) with voltage dividing resistor (R3) on the top (outward facing) surface of the sensing module. (\emph{b}) A reflective phototransistor (U1) is mounted on the bottom surace of the module, facing inwards towards the silicone body of the haptic device. An initial air-gap of $0.5 mm$ is maintained between the phototransistor and the device's silicone body using 3M VHB 4905 double-sided tap as a spacer. When pressure is applied to a module during contact with a physical surface, this airgap is closed and the phototransitor changes state, signaling contact. (\emph{c}) A $10 mm$ thermally conductive silicone pad is mounted on top of the temperature sensor to aid in even heat transfer and accurate temperature measurement of contact surfaces.} }
    \label{fig:Contact-Temp}
\end{figure}

The thermal display control loop we developed consist of a software PID
controller, the ADN8833 TEM driver IC, a differential filter with a cutoff frequency of 5 kHz, and a
thermistor mounted on the TEM's cold side for temperature feedback. \hl{Activation and temperature setpoints of TEMS during active temperature passthrough were controlled for each individual TEM by a corresponding surface contact and temperature measurement module, shown on the device in Figure \ref{fig:Ch4-DevicePics} and further detailed in Figure \ref{fig:Contact-Temp}.} While the drive circuit, filter,
and temperature and contact sensors are separate channels and contain unique hardware for each TEM, a single microcontroller is sufficient to run all the PID controllers and interface with a PC to receive thermal rendering input. Additional components of the control PCB include separate multiplexers for the thermistor readings, TEM current sense, and TEM voltage sense, allowing each to be read by the microcontroller using a single input pin each.

\subsection{Water Cooling} \label{ch4:WaterCooling}

The fact that a TEM acts as a resistive load leads to a large amount of excess heat generation. In fact, the resistive heating dumped on the cold side ($\frac{1}{2}I^2R$) can sometimes approach $Q_{max}$, the maximum heat transfer capable by the TEM. Additionally, as described in equation \ref{eq:ch4-temheatflow}, when in cooling mode natural thermal conduction within the TEM transfer heat from the hot side to the cold side. If not addressed, this heat buildup will quickly counter any attempts to lower the temperature of the TEM's cold side. To address this issue, we added aluminum heatsinks to the hot side of the TEM and implemented a novel watercooling solution to cool the heatsinks. Our solution is unique in that the water channels are built into the flexible and wearable silicone base itself, allowing for watercooling to be applied while still maintaining the mobile benefits of the wearable device already described. As seen in Figure \ref{fig:Ch4-DevicePics} (c-d), each TEM assembly is watercooled in series, allowing for steady volumetric flow across each TEM.

When implementing the watercooling solution, it was first necessary to determine the maximum total energy transfer that the water cooling solution would need to handle. This maximum necessary heat dissipation occurs when the TEM is in cooling mode, meaning the cold side is reduced in temperature and the hot side is increased in temperature. Flipping the signs of the $q_{pel}$ and thermal conduction components in equation \ref{eq:ch4-temheatflow}, we arrive at the following equation describing heat flow into the hot side:

\begin{equation} \label{eq:ch4-hotsideflow}
    q_{h} = \   \alpha T_{c}I + \frac{1}{R_{\text{th}}}\left( T_{c} - T_{h} \right) + \frac{1}{2}R_{\text{el}}I^{2}
\end{equation}

As we are looking at the worst case scenario, we assume that $T_c = T_h$ and the thermal conduction component goes away. The first component, representing $q_{pel}$, can be set as the $Q_{max}$ value of the specific TEM used. In our case, $Q_{max} = 1.7 W$. The maximum current for our TEM is $0.7A$, and its resistive load is $4.17 \Omega$. As there are nine total TEMs in our assembly, $Q_{max_{Array}}$ can be calculated as:

\begin{equation} \label{eq:ch4-Qmax}
    Q_{max_{Array}} = 9 * (1.7 + \frac{1}{2}4.17*0.7^2) = 24.49 W
\end{equation}

\noindent The maximum temperature rise for the water in the watercooling system can then be described by:

\begin{equation} \label{eq:ch4-DeltaTEq}
    \Delta T = \frac{Q_{max_{Array}}}{\rho Vc}
\end{equation}

\noindent where $\rho$ is the density of water ($1000 \frac{kg}{m^3}$), $V$ is the volumetric flowrate (measured to be $2.25\frac{ml}{s}$), and $c$ is the specific heat of water ($4182 \frac{J}{kg \cdot K}$). Plugging these values in we get a maximum temperature differential of:

\begin{equation} \label{eq:ch4-DeltaTValues}
    \Delta T = \frac{24.49}{1000*2.25E^{-6}*4184} = 2.60 K
\end{equation}

\noindent The intermittent use of the TEMS, combined with relatively small $\Delta T$ for the water and the relatively large size of our water reservoir, leads us to conclude that the water cooling system will remain stable even with prolonged use of the device.

A heating element within the water reservoir kept the water temperature at $30^{\circ}C$, which is within the range of skin temperature of $25^{\circ}C$ - $36^{\circ}C$ \cite{verrilloEffectsHydrationTactile1998,jonesMaterialDiscriminationThermal2003}. Due to the large size of the water cooling channels compared to the size of the device, and the thermal properties of the silicone material with which is was made, this maintained a steady temperature of $30^{\circ}C$ for the entire thermal device. Maintaining the wearable thermal device at skin temperature is of great importance for multiple reasons. First and foremost, doing so establishes a neutral baseline, enabling the user to perceive both warming and cooling stimuli from the device. This neutral baseline also minimizes sensory adaptation, thereby preserving the sensitivity and responsiveness of the user's skin to the thermal haptic feedback. Additionally, keeping the device at skin temperature when not actively providing thermal feedback enhances user comfort and reduces the likelihood of skin irritation or discomfort from prolonged contact with a device that is either too cold or too warm.

\subsection{Device Performance} \label{ch4:SystemPerformance}

\begin{figure}[htbp]
    \centering
    \includegraphics[width=\textwidth]{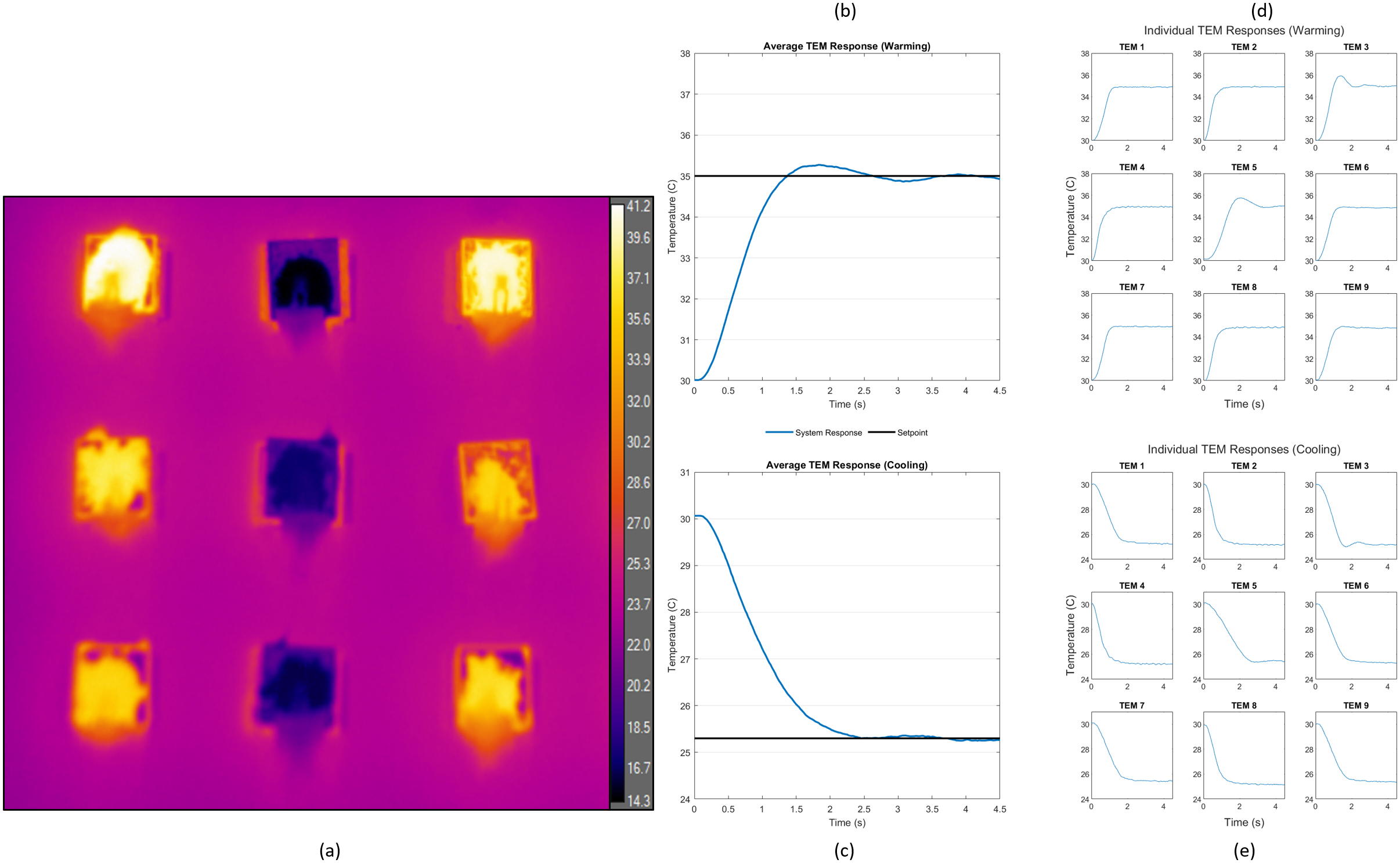}
    \caption{\textbf{Thermal Feedback Device Performance.}\\(\emph{a}) Each channel of the TEM array can be individually actuated, allowing for multiple spatial patterns to be formed and presented to the palm. (\emph{b-c}) Also shown in Figure \ref{fig:Ch4-DevicePics}(e). The average temperature response to a step input while in contact with a human palm for warming (b) and cooling (c) modes. (\emph{d-e}) Individual TEM responses for warming (d) and cooling (e) show that variance exists in an individual TEM's response, most likely due to variable contact pressure with the palm (and therefore variable rates of thermal transfer into the skin).}
    \label{fig:Ch4-TDFLIR-SystemResponse}
\end{figure}

The PID gains were tuned by hand, with the same gains applied to each TEM sub-system. Investigating the dynamic performance of the device, a step input exhibited a quick response for both warming and cooling modes while in contact with a human palm. A warming rise time of 1.4 seconds and a cooling rise time of 2.4 seconds were observed (Figure \ref{fig:Ch4-TDFLIR-SystemResponse} (b-e)). It is worth noting that the disparity in these rise times can be ascribed to the resistive heating of the TEMs, which facilitates the temperature increase in warming mode while impeding the temperature decrease in cooling mode. 

\FloatBarrier

\bibliography{references}

\section*{Acknowledgments}

We gratefully acknowledge the support provided for this research by the National Science Foundation under the award 2225890. We also thank the volunteers and the participants who agreed to be part of the experiments. The content is solely the responsibility of the authors and does not necessarily represent the official views of the National Science Foundation.

\section*{Author contributions statement}

A.W, R.G, and N.S conceptualized and designed the research. A.W., E.C., and R.G. designed and built the experimental hardware and conducted the experiment. A.W. and N.S analyzed the data. A.W. wrote the manuscript and prepared figures. All authors reviewed the manuscript.

\section*{Additional information}

\textbf{Competing interests} The authors declare no competing interests.

\textbf{Data Availability} Data is available at https://github.com/VU-RASL/ThermalFeedbackForAR-Data

\end{document}